\documentclass{article}
\usepackage{spconf,amsmath,graphicx,hyperref}
\usepackage{booktabs,tabularx}
\usepackage[table]{xcolor}
\usepackage{float}  
\usepackage{stfloats}
\usepackage{enumitem}
\usepackage{xspace}
\usepackage{makecell}
\usepackage{colortbl}

\newcommand{\tool}{\text{SP-MCQA}\xspace}

\title{\tool: Evaluating Intelligibility of TTS Beyond the Word Level}
\name{Hitomi Jin Ling Tee  \qquad Chaoren Wang  \qquad Zijie Zhang  \qquad Zhizheng Wu}
\address{The Chinese University of Hong Kong, Shenzhen}
\begin{document}
%\ninept
%
\maketitle
\begin{abstract}
The evaluation of intelligibility for TTS has reached a bottleneck, as existing assessments heavily rely on word-by-word accuracy metrics such as WER, which fail to capture the complexity of real-world speech or reflect human comprehension needs. To address this, we propose \tool (Spoken-Passage Multiple-Choice Question Answering), a novel subjective approach evaluating the accuracy of key information in synthesized speech, and release \tool-Eval, an 8.76-hour news-style benchmark dataset for \tool evaluation. Our experiments reveal that low WER does not necessarily guarantee high key-information accuracy, exposing a gap between traditional metrics and practical intelligibility. \tool shows that even state-of-the-art (SOTA) models still lack robust text normalization and phonetic accuracy. This work underscores the urgent need for high-level, more life-like evaluation criteria now that many systems already excel at WER yet may fall short on real-world intelligibility.
\end{abstract}
\begin{keywords}
TTS evaluation, subjective metric, key information accuracy, benchmark dataset
\end{keywords}
\section{Introduction}
\label{sec:intro}
Text-to-Speech (TTS) \cite{peng-etal-2024-voicecraft, anastassiou2024seed, 10.5555/3692070.3692979, le2024voicebox, liao2024fish, chen2024f5, du2024cosyvoice, wang2025maskgct} systems have achieved remarkable progress in producing highly intelligible speech. However, evaluation methods have not kept pace with these advances. Existing intelligibility-related metrics, such as Word Error Rate (WER) or subjective intelligibility Mean Opinion Score (MOS), predominantly focus on low-level accuracy which overlook if the key information is accurately conveyed. This gap is critical because, in real-world scenarios, listeners care more about understanding essential information than perfect word-by-word reproduction. 

Meanwhile, many existing TTS evaluation test sets \cite{panayotov2015librispeech, zen2019libritts, chen2021gigaspeech} are relatively simple and standardized, lacking challenging cases that reflect real-world speech variability. Although Seed-TTS \cite{anastassiou2024seed} introduces challenging patterns such as word repetitions and tongue twisters, it still fails to adequately assess model performance on irregular text — particularly content involving digits and proper nouns, such as locations, names, numbers, and events — which frequently constitute key information in informative, context-rich speech.

To address these shortcomings, we introduce \tool (Spoken-Passage Multiple-Choice Question Answering), a framework for evaluating TTS systems on key-information accuracy beyond the word level, comprising a novel subjective evaluation metric (\tool ACC) and test set (\tool-Eval). In line with the need for \tool evaluation, \tool-Eval is constructed as a news-style test set for speech synthesis that is both acoustically natural and semantically rich in contextual and critical information. Unlike traditional transcript-based metrics, \tool does not measure word-by-word accuracy and instead evaluates semantic and structural fidelity of key information under realistic listening conditions. It acts as a complementary framework to WER for evaluating and comparing models that already achieve high intelligibility at the word level. 

The contributions of our work are threefold:
\begin{itemize}[itemsep=1pt,topsep=0pt,parsep=0pt]
    \item We propose \tool, a novel subjective evaluation approach to measure the key information accuracy of a synthesized speech.
    \item We create \tool-Eval, a new open-source news-style benchmark dataset that contains uncommon text, involving proper nouns and digits, designed for \tool.
    \item We conduct an in-depth and comprehensive evaluation of how state-of-the-art (SOTA) TTS systems perform on this benchmark.
\end{itemize}

\begin{figure*}[ht]
    \vspace{-70pt}
    \includegraphics[width=1.0\linewidth,keepaspectratio]{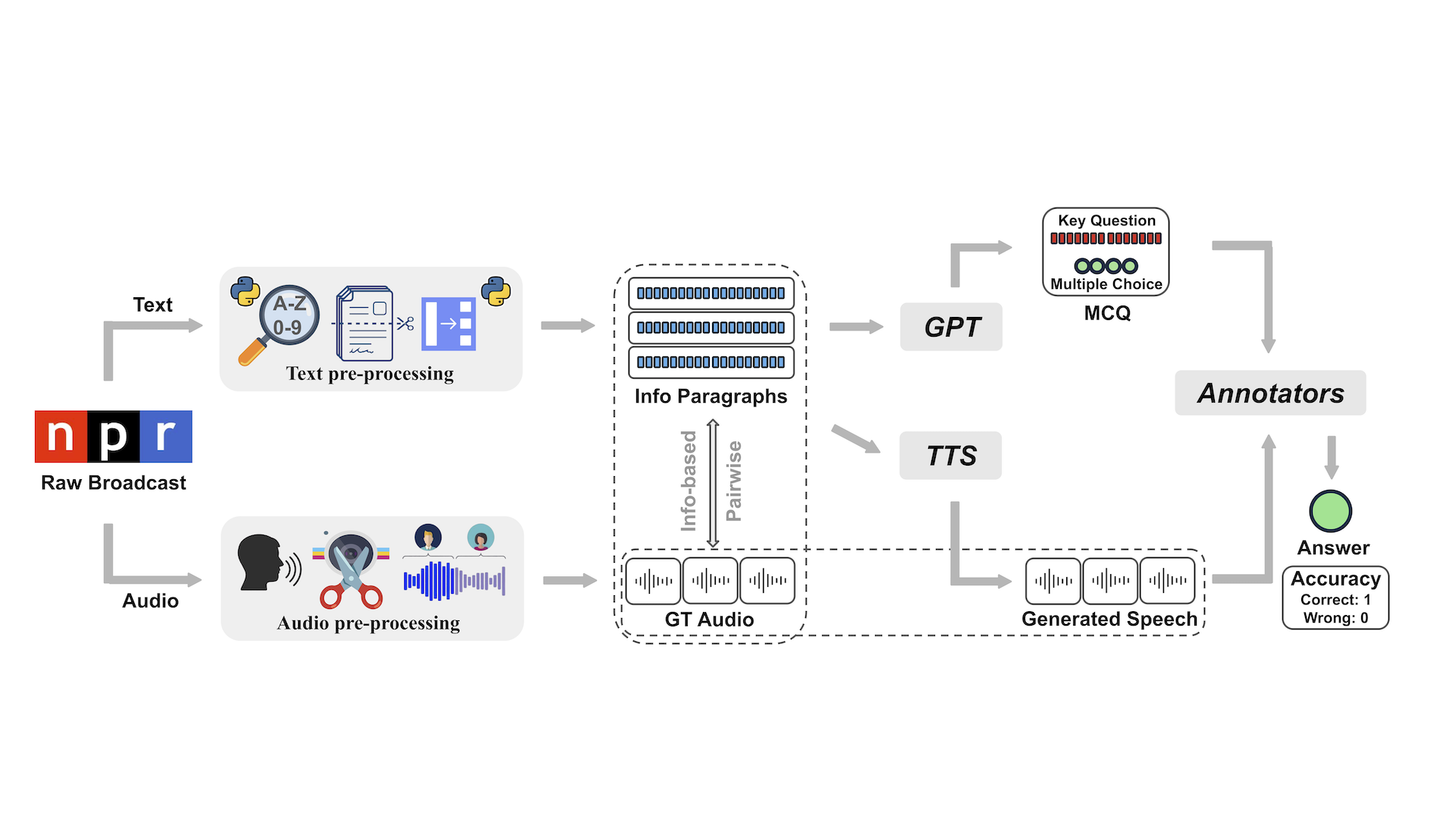}
    \vspace{-70pt}
    \caption{An overview of \tool. The process involves two main stages: (1) The creation of \tool-Eval benchmark dataset, and (2) the pipeline for \tool evaluation.}
    \label{fig:SMCQA}
    \vspace{-5pt}
\end{figure*}

\section{Related Work}
\label{sec:review}
Spoken Multiple-Choice Question Answering (SMCQA), a form of Spoken Question Answering (SQA), referred to tasks related to machine text comprehension in which passages, questions, and multiple choices are presented entirely in speech \cite{9003966, djeffal2023automatic}. In contrast, we study a hybrid setting similar to the text-based question answering on listening comprehension test \cite{raina2023analyzingmultiplechoicereadinglistening}, where only the passage is spoken while the questions and answer choices are in text. We term this new variant \tool (Spoken-Passage Multiple-Choice Question Answering) to clearly distinguish it from SMCQA.

There are several datasets for listening comprehension research, such as TOEFL-QA\footnote{\href{https://github.com/iamyuanchung/TOEFL-QA}{https://github.com/iamyuanchung/TOEFL-QA}} and TED-Q \cite{westera-etal-2020-ted}. Although these corpora resemble our newly created test set in form, they rarely focus on key information involving digits or numbers, making them less suitable than \tool-Eval for assessing the intelligibility of synthetic speech on key information.

\section{\tool Overview}
\label{sec:smcqa}

\subsection{Benchmark Dataset}
\label{sec:data_stats}
The \tool-Eval dataset contains proper nouns and digits (dates, names, times, locations, events, etc.). Table \ref{tab1} summarizes the \tool-Eval statistics, comprising 5,805 utterances with a total duration of 8.76 hours of speech.

\begin{table}[h]\small
    \vspace{-10pt}
    \centering
    \caption{Statistics of the \tool-Eval test set.}
    \begin{tabular}{cccccc}
        \toprule
       \rowcolor{gray!20} \textbf{\# Hrs} & \textbf{\# Spk} & \textbf{\# Para} & \textbf{\# Utts} & \textbf{\# Ques}
        \\
        \midrule
        8.76 & 483 & 550 & 5,805 & 2,688 \\
        \bottomrule
    \end{tabular}
    \label{tab1}
\end{table}

Audio and manually annotated text are sourced from National Public Radio (NPR) news, providing conversational speech with rich contextual information that facilitates the extraction of key details commonly encountered in daily life. 

Following Emilia’s pre-processing pipeline \cite{he2024emilia}, background music is removed using Ultimate Vocal Remover (UVR)\footnote{\href{https://github.com/Anjok07/ultimatevocalremovergui}{https://github.com/Anjok07/ultimatevocalremovergui}}
, and WhisperX \cite{bain2023whisperx} is used for ASR and timestamp extraction. Because uppercase letters and numbers often indicate key information, we apply Regular Expressions (RegEx) to filter the raw text, retaining paragraphs that contain: (1)~at least one number with a minimum of three digits; and (2)~at least two uppercase letters (excluding those at sentence beginnings). From this filtered pool, we randomly select 550 ``information paragraphs,'' each constrained to 65--260 words (\(\approx 30\,\text{s}-2\,\text{min}\) of speech). Then, Pydub is used to segment the clean audio according to these paragraphs, producing the ground-truth for subsequent \tool testing. We further apply pyannote speaker-diarization-3.1 \cite{bredin2023pyannote} to extract speaker timestamps and refine the segmentation to separate different speakers. Since most TTS systems are not trained on very long utterances, we use the Natural Language Toolkit (NLTK) \cite{loper2002nltk} to split each paragraph into natural sentences and segment the audio accordingly. These aligned sentence–audio pairs form the ground-truth for later objective evaluation.

We employ GPT-4o-mini \cite{hurst2024gpt} to automatically generate multiple-choice questions (MCQs) that examine the key information in each paragraph. Each evaluation task consists of a paragraph and its two to ten associated questions. Each MCQ contains four options: one correct answer, one “Other,” and two distractors representing different error types, as illustrated in Table \ref{tab2}. All questions are manually inspected, and problematic items caused by GPT hallucinations are removed.

\begin{table*}[htbp]
\vspace{-10pt}
\centering
\small
\caption{Error type descriptions of the generated multiple-choice questions.}
\begin{tabularx}{\textwidth}{lXl}
\toprule
\rowcolor{gray!20} \textbf{Error Type} & \textbf{Description} & \textbf{Example} \\
\midrule
Phonetic Error & Similar sounding to the correct answer. & “Eighteen” vs. “Eighty” \\
Semantic Error & Logically reasonable but factually incorrect. & “Wednesday” vs. “Thursday” \\
Syntax Error & Structural mistakes in phrasing. & “Ph.D. Emily Clark” vs. “Emily Clark, Ph.D.”  \\
Grammar Error & Grammatical inconsistencies or subtle inaccuracies in expression. & “Wet cloths” vs. “Cloths wet” \\
Other & Always appear as an option. & “None of the above” \\
\bottomrule
\label{tab2}
\vspace{-15pt}
\end{tabularx}
\end{table*}

\subsection{Evaluation Pipeline}
\label{ssec:data_stats}
For \tool evaluation, annotators listen to speech and answer multiple-choice textual questions based on its content, with accuracy scored as 1 for a correct answer and 0 otherwise. We recruit 40 annotators who are either native English speakers or non-native with an IELTS listening score 8.0\footnote{\href{https://takeielts.britishcouncil.org/teach-ielts/test-information/ielts-scores-explained}{https://takeielts.britishcouncil.org/teach-ielts/test-information/ielts-scores-explained} Band 8.0 indicates a very good user with a fully operational command of the language, making only occasional unsystematic inaccuracies or inappropriacies, and able to handle complex, detailed argumentation with only minor misunderstandings in unfamiliar situations.} and above. All annotators receive clear instructions: “Select the answer that directly matches the information explicitly stated in the audio; do not infer beyond what is clearly stated.” 

Each evaluation task is randomly assigned to two annotators. If their answers differ, a third annotator is added; if all three differ, a fourth is introduced. No further annotators are involved if disagreement persists among four. Golden test questions are randomly inserted into 10\% of tasks to assess general knowledge. Only annotators achieving 100\% accuracy on golden tests are retained; results from those below this threshold are discarded, and their tasks are reassigned to other qualified annotators. The final \tool ACC is computed as the average accuracy across qualified annotators. We also collect qualitative feedback for future analysis.

\section{Experiments}
\label{sec:experiments}

\subsection{Models}
\label{ssec:model}
We select FishSpeech V1.4 \cite{liao2024fish}, MaskGCT \cite{wang2025maskgct}, F5-TTS \cite{chen2024f5}, and CosyVoice 2 \cite{du2024cosyvoice} for our \tool evaluation due to their strong performance in regular intelligibility assessment. FishSpeech V1.4 employs a dual autoregressive (AR) architecture and leverages LLMs for rich linguistic feature extraction, resulting in clear pronunciation and highly intelligible output. MaskGCT, a fully non-autoregressive (NAR) model with a mask-and-predict paradigm, enables fast parallel synthesis while maintaining high word-level accuracy and strong speaker similarity across unseen speakers. F5-TTS, also fully NAR with flow-matching-based diffusion, achieves low WER and expressive speech synthesis, providing robust intelligibility and controllable speaking rates. CosyVoice 2 combines AR fidelity with NAR speed, separately modeling semantic and acoustic features to enhance prosody, rhythm, and intonation, supporting naturalness and speaker identity preservation in both streaming and non-streaming scenarios. Collectively, these models represent diverse architectures and design choices, making them ideal for assessing the accuracy, speaker consistency, and audio quality of synthesized speech in key-information-sensitive tasks under news-style content.

\subsection{Evaluation Metrics}
\label{ssec:metric}
We also conduct objective evaluations on each model across three key aspects: intelligibility, coherence, and audio quality. Intelligibility is evaluated using Word Error Rate (WER), computed with the jewel package and transcribed by Whisper-large-v3\footnote{\href{https://huggingface.co/openai/whisper-large-v3}{https://huggingface.co/openai/whisper-large-v3}} ASR, with a fixed ``prompt'' parameter to ensure consistent transcription style. Coherence is assessed via speaker similarity (S-SIM), which computes the cosine similarity between WavLM-TDNN \cite{chen2022wavlm} speaker embeddings of the synthesized speech and the reference prompt. Audio quality is evaluated using Deep Noise Suppression Mean Opinion Score (DNSMOS) \cite{reddy2022dnsmos}, derived from P.835 human ratings, providing an overall audio quality score on a 1-5 scale. All metrics are computed at a 16 kHz sampling rate. While WER is our primary intelligibility metric for comparison, all three objective metrics together provide reference values for the overall quality of the \tool-Eval test set and the synthesized speech produced by TTS models.

\subsection{Experimental Setup}
\label{ssec:setup}
The SP-MCQA-Eval test set contains 483 unique speakers. During inference, we select one prompt (utterance + transcript) from each speaker, while the transcripts of all utterances—including the selected prompts—serve as target texts, yielding 5,805 prompt–target pairs in total. These pairs preserve speaker identity, enabling direct evaluation against ground-truth data. TTS inferences are conducted on 8 NVIDIA GeForce RTX 4090 GPUs using the official code from each model’s GitHub repository. For MaskGCT\footnote{\href{https://github.com/open-mmlab/Amphion/tree/main/models/tts/maskgct}{https://github.com/open-mmlab/Amphion/tree/main/models/tts/maskgct}}, we modify the G2P module to correctly classify numerical inputs as English (en) instead of ``other language.'' For CosyVoice 2\footnote{\href{https://github.com/FunAudioLLM/CosyVoice}{https://github.com/FunAudioLLM/CosyVoice}}, we remove the duration constraint in frontend.py, allowing the model to process speech tokens for audio longer than 30 seconds. No modifications are made to the inference code for FishSpeech V1.4\footnote{\href{https://github.com/fishaudio/fish-speech}{https://github.com/fishaudio/fish-speech}} and F5-TTS\footnote{\href{https://github.com/SWivid/F5-TTS}{https://github.com/SWivid/F5-TTS}}. All evaluations are also conducted under the same experimental setup as inference.

\subsection{Results \& Analysis}
\begin{table}[h]\small
\vspace{-20pt}
\centering
\caption{Evaluation results of ground-truth and four TTS models on \tool-Eval test set.}
\setlength{\tabcolsep}{4pt} 
\renewcommand{\arraystretch}{1.1} 
\begin{tabular}{ccccc}
\toprule
\rowcolor{gray!20} 
\textbf{System} &
\textbf{\makecell{\tool\\ ACC (\%) $\uparrow$}} &
\textbf{\makecell{WER\\ (\%) $\downarrow$}} &
\textbf{S-SIM $\uparrow$} &
\textbf{\makecell{DNSMOS\\P.835 OVRL $\uparrow$}} \\
\midrule
Ground-Truth & 92.045 & 8.067 & 0.710 & 2.955 \\
F5-TTS       & 87.139 & 11.267 & 0.654 & 3.202 \\
MaskGCT      & 89.260 & 7.351  & \textbf{0.710} & 3.081 \\
CosyVoice 2  & \textbf{90.399} & 9.044  & 0.523 & \textbf{3.334} \\
FishSpeech   & 81.194 & \textbf{5.739} & 0.522 & 3.242 \\
\bottomrule
\end{tabular}
\label{tab3}
\vspace{-5pt}
\end{table}

\begin{table*}[!t]
\small
\vspace{-10pt}
\centering
\caption{Analysis of error types on \tool evaluation for each system.}
\begin{tabular}{lcc|cccc}
\toprule
\rowcolor{gray!20} 
\textbf{System} & \multicolumn{2}{c|}{\textbf{Evaluation}} & \multicolumn{4}{c}{\textbf{Error Types}} \\
    \cmidrule(lr){2-3} \cmidrule(lr){4-7}
 & \textbf{Total Ques} & \textbf{Wrong Ques} & \textbf{Phonetic} & \textbf{Semantic} & \textbf{Structure (Syntax + Grammar)} & \textbf{Other} \\
\midrule
Ground-Truth & 6914 & 550 & 246 (\textbf{3.558\%}) & 80 (1.157\%) & 49 + 61 (1.591\%) & 114 (1.649\%) \\
F5-TTS       & 7472 & 961 & 306 (\textbf{4.095\%}) & 114 (1.526\%) & 79 + 93 (2.302\%) & 369 (4.938\%) \\
MaskGCT & 7477 & 803 & 267 (\textbf{3.571\%}) & 104 (1.391\%) & 74 + 93 (2.234\%) & 265 (3.544\%) \\
CosyVoice 2   & 7218 & 693 & 233 (\textbf{3.228\%}) & 70 (0.970\%) & 64 + 72 (1.884\%) & 254 (3.519\%) \\
FishSpeech   & 7519 & 1414 & 271 (3.604\%) & 104 (1.383\%) & 66 + 77 (1.902\%) & 896 (\textbf{11.916\%}) \\
\bottomrule
\end{tabular}
\label{tab5}
\vspace{-5pt}
\end{table*}

\begin{table*}[!t]
\centering
\small
\caption{Main comments for selecting “Other” in \tool evaluation (lightly edited for clarity).}
\begin{tabularx}{\textwidth}{ccXc}
\toprule
\rowcolor{gray!20}
\textbf{Task ID} & \textbf{System} & \textbf{Comment} & \textbf{Related Issue} \\
\midrule
2210 & CosyVoice 2 & A “-nine” sound occurs after every sentence. (Cannot recognize). & Noise \\
1938 & CosyVoice 2 & Each sentence ends with an “-edge”/“‑ged” sound, causing confusion. (Cannot recognize). & Noise \\
951  & MaskGCT    & The audio is noisy and difficult to identify. (Cannot recognize). & Noise \\
975  & MaskGCT    & In the first 55 seconds, pronunciation was unclear, resembling reversed speech or persistent difficulty articulating the governor’s name throughout the segment. (Cannot recognize). & Noise \\
543  & MaskGCT    & Pronunciation resembles “Alala” rather than “Alabama.” (No options). & Proper Noun \\
1156 & F5-TTS     & Too fast with unclear pronunciation. (Cannot recognize). & Speed \\
380  & F5-TTS     & Speech rate is approximately 1.75×. (Cannot recognize). & Speed \\
544  & F5-TTS     & Pronunciation alternates between “... Ala” and “... Alabama.” (Inconsistency). & Proper Noun \\
689  & FishSpeech & The number is 2, not 2,000.	(No options). & Number \\
541  & FishSpeech & Only “Talladega” is audible, not “Talladega Ala” or “Talladega Alabama.” (No options). & Proper Noun \\
185  & FishSpeech & No data was mentioned. (No options). & Number \\
\bottomrule
\label{tab4}
\vspace{-10pt}
\end{tabularx}
\end{table*} 

The evaluation results of ground-truth and four TTS models under \tool-Eval are shown in Table \ref{tab3}. The relatively moderate performance of the ground-truth mainly stem from imprecise in timestamp extraction during pre-processing; nevertheless, these values still provide a meaningful reference for evaluation. While FishSpeech ranks highest in WER, it performs worst in \tool ACC, whereas CosyVoice 2, despite its lower rank in WER, ranks highest in \tool ACC. This reveals a critical limitation of WER: models with low WER may still fail to convey key information accurately.

Table~\ref{tab5} presents the analysis of the error types, showing that phonetic errors are the most prevalent across all models, followed by structural (syntax and grammar) and semantic errors. The latter two might be caused by model's “hallucinations,” such as generating inconsistent content with the input, or omitting content. We observe that NAR models (F5-TTS and MaskGCT) exhibit a higher proportion of such errors compared to AR models (CosyVoice2 and FishSpeech). Nonetheless, regardless of architecture, phonetic accuracy remains the primary challenge in our \tool tasks for key information, likely due to the scarcity of irregular or uncommon patterns in training data. These findings highlight the importance of addressing phonetic errors, especially in rare or atypical utterances, for developing human-like TTS systems. 

We further analyze annotators' comments for selecting “Other” (Table \ref{tab4}). CosyVoice 2 occasionally adds noise at sentence endings, raising WER and slightly lowering \tool ACC. However, it is the only model that correctly pronounces all tested abbreviations (e.g., “Ala.”→“Alabama”), achieving the highest \tool ACC. MaskGCT generates odd sounds and struggles with proper nouns (e.g., “Ala.”→“Alala”), performing slightly worse on the \tool task. F5-TTS, in addition to incorrectly handling uncommon text (e.g., “Ala.”→“Ala”), exhibits overly fast speech, hindering the recognition of key information. FishSpeech suffers from mid-sentence word drops and cut-offs due to normalization issues (e.g., “Ala.”→“ ”), yielding the lowest \tool ACC despite lowest WER. These observations underscore the need for robust text normalization and key-information accuracy evaluation beyond conventional metrics.

\section{Conclusion}
\label{sec:conclusion}
In this work, we propose \tool, a framework that evaluates the accuracy of key information in synthesized speech beyond the word level for TTS models. \tool demonstrates that even SOTA TTS models exhibit critical weaknesses in key information when handling real-world speech complexity. Key findings include: (1) significant discrepancies between word-by-word accuracy and key information accuracy; (2) phonetic errors and text-normalization challenges in uncommon contexts and irregular patterns — particularly with names, numbers, and abbreviations — with each model exhibiting distinct error patterns, all of which should be addressed in future speech synthesis research. While our approach paves the way for high-level evaluation, limitations such as the substantial manual effort required for human evaluation remain. Future work will explore leveraging Audio LLMs for more efficient and scalable assessment and extend this work to other languages to provide clearer guidance for improving multilingual TTS systems.

\clearpage
\clearpage

\bibliographystyle{IEEEbib}
\bibliography{main}

\end{document}